\documentclass[aps,prl,twocolumn,groupedaddress]{revtex4}
\usepackage{amssymb,graphicx,xspace,color}
\newcommand{\tc}{$T_c$\xspace}
\newcommand{\tn}{$T_N$\xspace}
\newcommand{\ts}{$T_S$\xspace}
\newcommand{\laf}{LaFeAsO$\mathrm{_{1-x}}$F$\mathrm{_{x}}$\xspace}
\newcommand{\lao}{LaFeAsO\xspace}

\newcommand{\smf}{SmFeAsO$\mathrm{_{1-x}}$F$\mathrm{_{x}}$\xspace}

\newcommand{\bfca}{BaFe$\mathrm{_{2-x}}$Co$\mathrm{_x}$As$\mathrm{_2}$\xspace}

\newcommand{\cfca}{CaFe$\mathrm{_{2-x}}$Co$\mathrm{_x}$As$\mathrm{_2}$\xspace}

\begin{document}

\title{Unusual Nernst effect and spin density wave precursors in superconducting $\rm\bf LaFeAsO_{1-x}F_x$}


\author{Agnieszka Kondrat, G\"unter Behr, Bernd B\"uchner}
\affiliation{Leibniz-Institute for Solid State and Materials Research, IFW-Dresden, 01171 Dresden, Germany}
\author{Christian Hess}
\email[]{c.hess@ifw-dresden.de}
\affiliation{Leibniz-Institute for Solid State and Materials Research, IFW-Dresden, 01171 Dresden, Germany}

\date{\today}

\begin{abstract}
The Nernst effect has recently proven as a sensitive probe for detecting unusual normal state properties of unconventional superconductors. Here we present a systematic study of the Nernst effect of the iron pnictide superconductor \laf with a particular focus on its evolution upon doping. For the parent compound we observe a huge negative Nernst coefficient accompanied with a severe violation of the Sondheimer cancellation in the spin density wave (SDW) ordered state. Surprisingly, an unusual and enhanced Nernst signal is also found at underdoping ($x=0.05$) despite the presence of bulk superconductivity and the absence of static magnetic order, strongly suggestive of SDW precursors at $T\lesssim150$~K.  A more conventional normal state Nernst response is observed at optimal doping ($x=0.1$) where it is rather featureless with a more complete Sondheimer cancellation.
\end{abstract}


\maketitle

The Nernst effect of unconventional superconductors has recently attracted a lot of attention for several reasons \cite{Xu2000,Wang2001,Wang2006,Cyr-Choiniere2009,Hackl2009a,Hackl2010,Behnia2009,Daou2010,Hackl2009}. For type-II superconductors it is strongly enhanced by movement of magnetic flux lines (vortices) \cite{Huebener1969,Otter1966,Hagen1990,Ri1994}. Here the Nernst coefficient $\nu$ is directly proportional to the drift velocity of the vortices, which has rendered this transport quantity a valuable tool for studying their dynamics. Based on this very fundamental property, the unusual enhancement of the Nernst coefficient in the normal state of cuprate high \tc superconductors at temperatures much higher than the critical temperature $T_c$ has been interpreted as the signature of vortex fluctuations \cite{Xu2000,Wang2001,Wang2006}. More specifically, it was proposed that in the pseudogap phase above $T_c$ long-range phase coherence of the superconducting order parameter is lost while the pair amplitude remains finite.
As an alternative explanation it has only recently been proposed that Fermi surface distortions due to stripe or spin density wave (SDW) order could lead to an enhanced Nernst effect in the cuprates \cite{Cyr-Choiniere2009,Hackl2009a,Hackl2010}. Very recently, a strong anisotropy of the Nernst coefficient arising from the broken rotation symmetry of electron-nematic order has been discussed both experimentally and theoretically \cite{Daou2010,Hackl2009}.

The discovery of superconductivity in \laf \cite{Kamihara2008} initiated a tremendous research effort which yielded soon after a large variety of new superconducting iron pnictide compounds with \tc up to 55~K \cite{Ren2008c}. 
The parent compound \lao is a poor metal and exhibits an antiferromagnetic SDW ground state. The transition towards the SDW state occurs at $T_N=137$~K and is accompanied by a structural tetragonal-to-orthorhombic transition at $T_s\approx160$~K \cite{Cruz2008,Klauss2008,Kondrat2009}.
Upon substituting fluorine for oxygen the SDW phase is destabilized, i.e. $T_s$ and $T_N$ gradually decrease and at some finite doping level ($x\lesssim0.05$) superconductivity emerges. The actual nature of the doping-driven transition from SDW to superconductivity is much under debate. There is evidence that in \laf the transition is abrupt and first order-like towards a homogeneous superconducting state \cite{Luetkens2009} while in other systems (e.g. \smf or \bfca) experiments suggest a finite doping interval where superconductivity and static magnetism coexist \cite{Drew2009,Ni2008}. 
This obvious proximity to antiferromagnetism suggests spin fluctuations being important for the mechanism of superconductivity with a respective impact on the normal state properties, including the normal state transport \cite{Kondrat2009,Prelovsek2009,Prelovsek2010,Hess2009}.

However, little is known about the Nernst effect of the pnictide superconductors family. In a pioneering study Zhu et al. reported an anomalous suppression of the off-diagonal thermoelectric current in optimally doped \laf and suggested the presence of SDW fluctuations near the superconducting transition \cite{Zhu2008}. More recently, Matusiak et al. observed a strong enhancement of the Nernst coefficient in the SDW state of the parent compound of \cfca, but did not find any particular anomaly in the Nernst effect of a superconducting doping level that could be attributed to neither vortex flow nor to SDW fluctuations \cite{Matusiak2010}.

In this letter we systematically investigate the doping-evolution of the Nernst effect in \laf. For the parent compound we observe a huge negative Nernst coefficient accompanied with a severe violation of the so-called Sondheimer cancellation in the SDW state. Surprisingly, at underdoping ($x=0.05$) we find a similar enhanced $\nu$ at $T\lesssim150$~K despite the absence of static magnetic order and the presence of bulk superconductivity, strongly suggestive of SDW fluctuations. More conventional transport is observed at optimal doping ($x=0.1$) where the normal state Nernst signal is rather featureless with a more complete Sondheimer cancellation.

Polycrystalline samples of \laf ($x=0$, 0.05, 0.1) were prepared and characterized by powder X-ray diffraction (XRD) and wavelength-dispersive X-ray spectroscopy (WDX) \cite{Kondrat2009} and further investigated by resistivity \cite{Hess2009}, magnetization \cite{Klingeler2010}, nuclear quadrupole resonance (NQR) \cite{Lang2010}, and muon spin rotation ($\mu$SR) experiments \cite{Klauss2008,Luetkens2009}.
Measurements of the Nernst, Seebeck and Hall coefficients were performed on cuboid-shaped samples of size $\sim0.6\times0.6\times1.5\rm~mm^3$ in a home-made device. The Seebeck and Nernst coefficient were measured  using a steady-state method with an SMD resistor as a heater. The temperature gradient on the sample was measured using a Au-Chromel differential thermocouple \cite{Hess2003b}. The Nernst coefficient was first determined as a function of temperature $T$ at constant magnetic field $B=14~$T with two opposite field polarizations, to compensate the longitudinal thermal voltages and then subsequently measured at several temperature points as a function of magnetic field. Excellent agreement was obtained between the two methods. The Hall effect was measured using a standard four-point method by sweeping the magnetic field at stabilized temperatures.

\begin{figure}[t]
\includegraphics[clip,width=1\columnwidth]{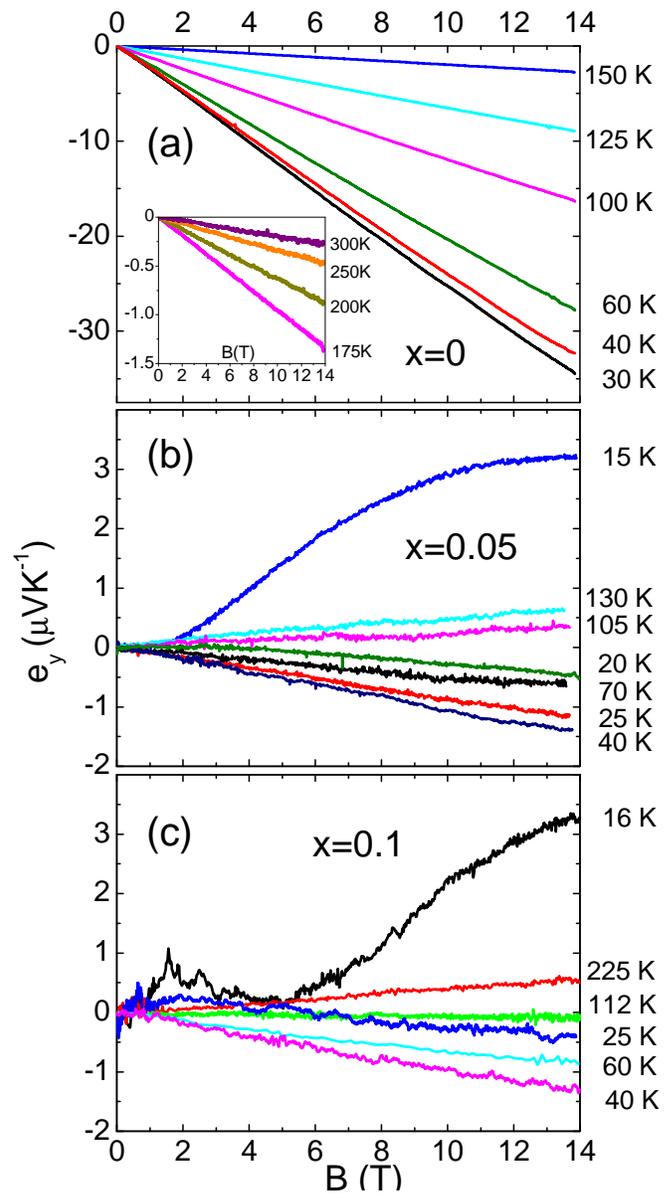}
\caption{Nernst signal $e_y$ of \laf as a function of magnetic field for $x=0$ (a), $x=0.05$ (b), $x=0.1$ (c).} 
\label{fig:figure1}
\end{figure}

Fig.~\ref{fig:figure1} (a) presents the field dependence of the Nernst signal $e_y=E_y/\nabla T$ (with $E_y$ the transverse electrical field) of the parent compound ($x=0$) for selected $T$. In the whole investigated temperature range $e_y$ is negative and within the experiment's accuracy linearly field dependent.
Between room temperature and 175~K the magnitude of the Nernst signal at 14~T is rather small with $|e_y|\lesssim1.3~\mu$V/K. 
This changes drastically upon crossing \ts and \tn where a strong enhancement of the Nernst signal occurs. For example, we find a strikingly large magnitude of $|e_y|\approx35~\mu$V/K at 30~K and 14~T.

Significantly different is the behavior of the Nernst coefficient of the underdoped compound at $x=0.05$ shown at Fig \ref{fig:figure1} (b). At 15~K we observe a positive signal with the amplitude 3~$\mu$V/K at 14T and a nonlinear magnetic field dependence which is indicative of strong contribution caused by vortex flow. As  $T$ rises, $e_y(B)$ changes sign and becomes linear just above $T_c=20.6$~K. With further increasing $T$ the Nernst signal reaches a minimum at 40~K and becomes positive again above ca. 100 K. The absolute values are more than one order of magnitude lower than those of LaFeAsO.

At optimal doping ($x=0.1$) the Nernst signal is at first glance very similar to that at $x=0.05$: a strongly non-linear and positive $e_y(B)$ due to vortex flow is observed below $T_c=26.8~$K. As $T$ rises, $e_y(B)$ becomes negative and linear  with a minimum at about 40~K, which is followed by another sign change at further elevated $T$. 

\begin{figure}[h]
\includegraphics[clip,width=1\columnwidth]{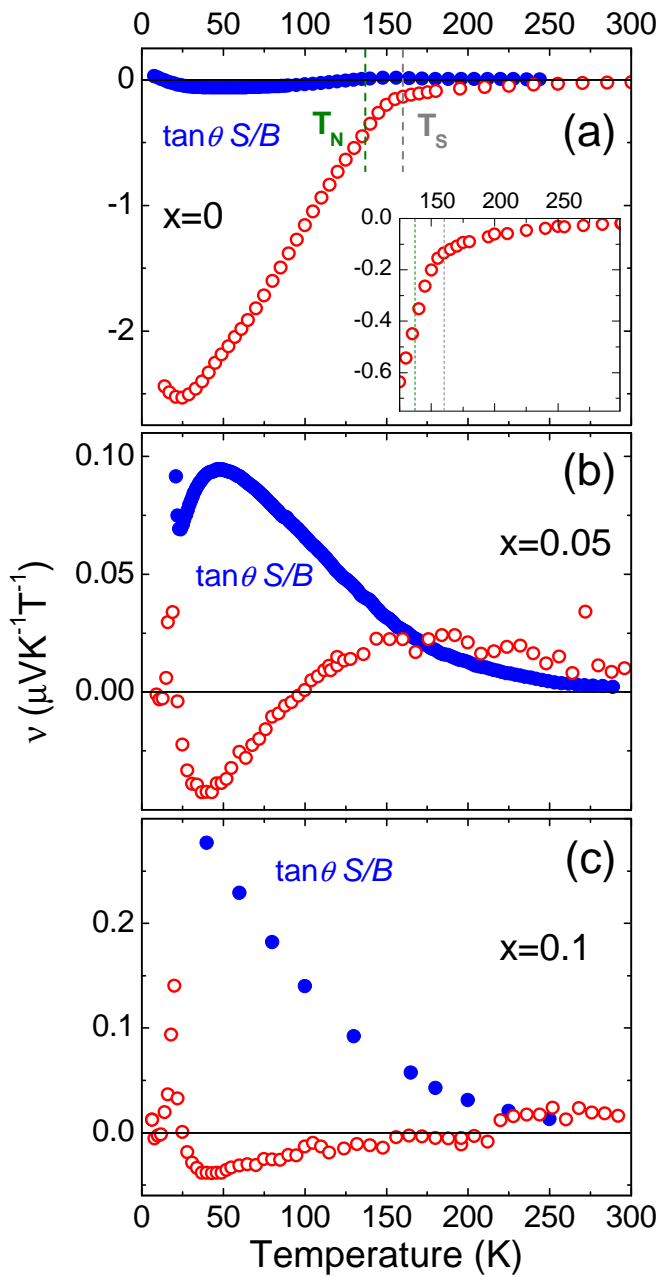}
\caption{Nernst coefficient $\nu$ (open circles) and $\tan\theta S/B$ (full circles) of \laf as a function of temperature for $x=0$ (a), $x=0.05$ (b), $x=0.1$ (c).} 
\label{fig:figure2}
\end{figure}

\begin{figure}[h]
\includegraphics[clip,width=1\columnwidth]{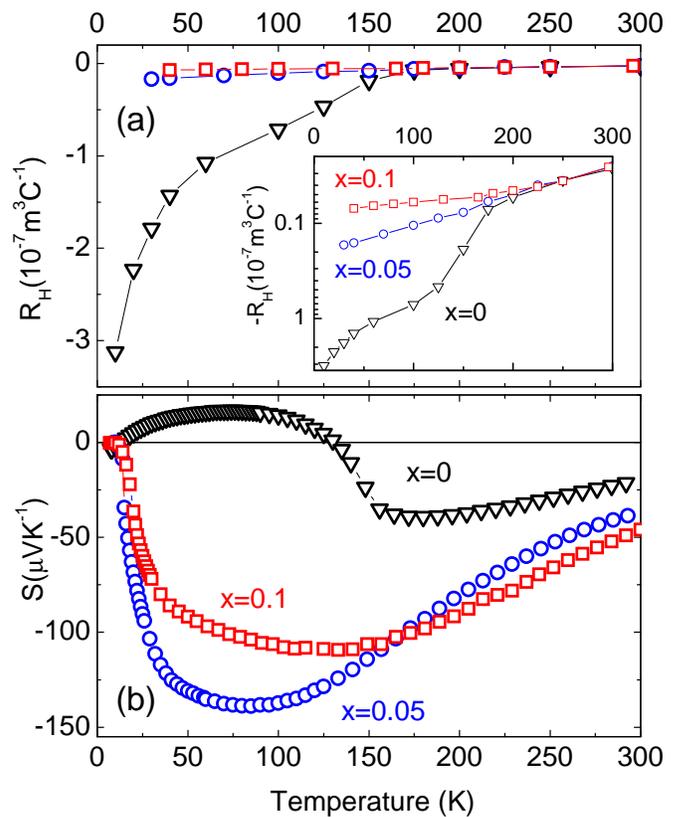}
\caption{Hall coefficient $R_H$ (a) and thermopower $S$ (b) as a function of temperature for $x=0$, 0.05, 0.1.}
\label{fig:figure3}
\end{figure}

We turn now to the actual $T$-dependence of the Nernst coefficient $\nu=e_y/B$ of the three samples shown in Fig.~\ref{fig:figure2}.
For the undoped sample, $\nu(T)$ is negative over the whole $T$ range. From $\nu=-0.02~\mu\rm VK^{-1}T^{-1}$ at 300~K the Nernst coefficient decreases moderately down to $\nu=-0.2~\mu\rm VK^{-1}T^{-1}$ at about 150~K. At $T\lesssim150$~K a large negative contribution is visible in the curve -- the slope changes strongly and $\nu(T)$ reaches its minimum at around 25~K with the 'giant' value of $-2.5~\mu\rm VK^{-1}T^{-1}$. Note that the magnitude of this Nernst response is about one order of magnitude larger than that generated by vortex flow in the superconducting samples or in, e.g., cuprate superconductors \cite{Wang2006,Ri1994} which is often considered as a benchmark for a large Nernst effect. Instead its size rather compares with the so-called 'giant' Nernst effect of $\rm URu_2Si_2$ and Bechgaard salts \cite{Bel2004,Nam2006,Behnia2009} \footnote{A qualitatively similar impact of SDW order on the Nernst effect has been observed also in CaFe$_2$As$_2$ by Matusiak et al. \cite{Matusiak2010}. However, the absolute magnitude of $\nu$ is almost one order of magnitude larger in our data for \lao.}.
In order to judge to what extent $\nu$ is unusual in terms of its magnitude, we consider the expression \cite{Wang2001,Sondheimer1948}
\begin{equation}
{\nu=(\frac{\alpha_{xy}}{\sigma}-S\tan \theta)\frac{1}{B}}. \label{sondheimer}
\end{equation}
Here $S$ is the Seebeck coefficient, $\tan \theta$ the Hall angle, $\sigma$ the electrical conductivity, and $\alpha_{xy}$ the off-diagonal Peltier conductivity. In a one-band metal, the two terms in Eq.~\ref{sondheimer} cancel exactly if the Hall angle is independent of energy ('Sondheimer cancellation') \cite{Sondheimer1948,Behnia2009,Wang2001}. However, in a multiband material such as \laf under scrutiny here, this cancellation is no longer valid. The degree of its violation can be determined experimentally by comparing the measured $\nu$ with the term $S\tan\theta/B$, which can be easily calculated from our thermopower and Hall data displayed in Fig.~\ref{fig:figure3} and the resistivity \cite{Hess2009}. The direct comparison of both quantities shown in Fig.~\ref{fig:figure2}a reveals clearly $|\nu|\ll |S\tan\theta|/B$, i.e. a \textit{severe} violation of the Sondheimer cancellation in the SDW phase.

The Nernst coefficient of the underdoped sample ($x=0.05$, Fig.~\ref{fig:figure2}b) exhibits a strong positive contribution arising from vortex motion in the superconducting state which extends up to a remarkably large $T\approx40$~K. Surprisingly, at higher temperature $\nu(T)$ resembles that of the SDW ordering as seen in the parent compound: Between 300~K and about 150~K we observe a rather flat $\nu(T)$ with a weak negative slope. However, at around 150~K the $T$-dependence changes and a sizeable negative contribution becomes apparent in $\nu$ which leads to a sign change at $\sim100$~K and a minimum at $\sim 40$~K where the vortex contribution sets in. Note that in this regime $|\nu|\approx |S\tan\theta|/B$, i.e. a significant violation of the Sondheimer cancellation is still obvious and that the negative contribution between 40~K and 150~K is of similar size as the vortex contribution at low $T$.
The presence of the SDW-like signature at $T\lesssim150$~K is completely unexpected since our sample exhibits bulk superconductivity where $\mu$SR and M\"ossbauer spectroscopy show no trace of magnetic ordering in this $T$-regime \cite{Luetkens2009}. We therefore conclude that instead of true SDW order, a form of a respective precursor, such as fluctuations or possibly nematic phases give rise to the enhanced Nernst response \cite{Hackl2009a,Hackl2009}. In fact, we observe an enhancement of $\nu$ with a similar magnitude also for the parent compound at $T\gtrsim T_N$ (c.f. Fig.~\ref{fig:figure1}a), i.e. in a $T$-range where SDW precursors are truly present \cite{Nakai2008,Wang2009a}, which corroborates this conclusion.

Despite a very similar behavior in the vicinity of \tc and similar magnitude the Nernst coefficient of the optimally doped compound ($x=0.1$, Fig.~\ref{fig:figure2}c) displays a completely different normal state behavior as compared to the underdoped compound. At $T\gtrsim40$~K, i.e. in the whole normal state, $\nu(T)$ is rather featureless with a weak positive slope.
In particular, no anomaly that could be related to SDW fluctuations is present and, moreover, a more complete Sondheimer cancellation, i.e. $|\nu|\gg |S\tan\theta|/B$ is found at low $T$. Altogether this suggests a more conventional transport behavior for this doping level.

Our findings provide fresh information about the doping evolution of the normal state properties of \laf, in particular concerning the interesting nature of the normal state close to the boundary to the SDW phase. 
We point out that the onset of anomalous Nernst response at $T\lesssim150$~K for $x=0.05$ coincides with a resistivity anomaly at about the same temperature which has been interpreted as a signature of SDW fluctuations as well \cite{Hess2009}. Furthermore, recent NMR and NQR studies provide clear-cut evidence for the slow-down of spin fluctuations in this $T$-regime \cite{Hessgrafeunpub} and for nanoscale electronic order \cite{Lang2010}, which further underpins our conclusion.
It thus seems that in contrast to other pnictide systems such as \bfca where strong evidence for the coexistence of static magnetic order and superconductivity exists \cite{Laplace2009}, \laf represents one rare example among the iron pnictides where both order parameters \textit{do not} coexist but truly compete with each other. However, one might speculate that rather small perturbations of the system, e.g. the disorder that Co-doping generates in the $\rm Fe_2As_2$ plane in \bfca, might be sufficient to stabilize the ubiquitous SDW fluctuations and drive the system towards the coexistence of static magnetism and superconductivity \cite{Fernandes2010,Fernandes2010a}.

It seems noteworthy that our results provide one more example of an unconventional superconductor where an unusually large normal state Nernst effect is observed. The Nernst signal related  to the SDW precursors clearly exceeds that of recently reported stripe and/or nematic phases observed recently in cuprate superconductors \cite{Cyr-Choiniere2009,Daou2010}. Furthermore, the negative sign of the unusual contribution to the Nernst coefficient unambiguously rules out vortex fluctuations \cite{Wang2006,Wang2001,Xu2000} as a thinkable origin since these should give rise to a positive Nernst response.

In summary, an unusually enhanced Nernst effect accompanied with a considerable violation of the Sondheimer cancellation is observed in both undoped as well as in underdoped \laf at $T\lesssim150$~K. 
While the enhancement in the parent compound is huge ('giant' Nernst effect) and coincides with the onset of SDW order, it is about one order of magnitude smaller in the underdoped, bulk superconducting sample. Since no trace of magnetic order is observed, our results suggest the presence of SDW fluctuations at this doping level. At optimal doping a more conventional, rather featureless Nernst response with a more complete Sondheimer cancellation is observed.

This work was supported by the Deutsche Forschungsgemeinschaft through Grant No. BE1749/12, the Research Unit FOR538 (Grant No.BU887/4) and the Priority Programme SPP1458 (Grant No. GR3330/2).

\end{document}